\documentclass[12pt]{article}
\usepackage{a4wide}
\usepackage{cite}
\usepackage{latexsym}
\usepackage{epsfig}
\usepackage{amsmath}
\usepackage{amssymb}

\usepackage{graphicx}
\usepackage{caption}
\usepackage{subcaption}

\usepackage{slashed}
\usepackage{color}

\begin{document}
 \date{ }
\title{ \bf Estimating VaR in credit risk: Aggregate vs single loss distribution}
\author{  \sc M. Assadsolimani\thanks{Email: m.assadsolimani@gmail.com}\,  and  D. Chetalova\thanks{Email: dchetalova@gmail.com} }
 \maketitle

\begin{abstract}\noindent
Using Monte Carlo simulation to calculate the Value at Risk (VaR) as a possible risk measure requires 
adequate techniques. One of these techniques is the application of a compound distribution for the aggregates in a portfolio. In this paper, we consider the aggregated loss of Gamma distributed severities and estimate the VaR by introducing a new approach to calculate the quantile function of the Gamma distribution at high confidence levels.  We then compare the VaR obtained from the aggregation process with the VaR obtained from a single loss distribution where the severities are drawn first  from an exponential and then from a truncated exponential distribution. We observe that the truncated exponential distribution as a model for the severities yields results closer to those obtained from the aggregation process. The deviations depend strongly on the number of obligors in the portfolio, but also on the amount of gross loss which truncates the exponential distribution.  
\end{abstract}

\newpage

\section{Introduction}\label{sec:introduction}
 
One main challenge of credit risk management is to estimate the loss distribution for credit portfolios. The loss distribution depends on the distribution of defaults within the portfolio and on the losses associated with each default.
Given a loss distribution, the Value at Risk (VaR) is a widely used measure to calculate the risk of loss. There are various methods to estimate the loss distribution and to compute the VaR. Monte Carlo simulation is one of these methods~\cite{opac-b1131339}. The estimation of VaR demands adequate quantification techniques especially if the portfolio is rather large. One of these techniques is the application of a compound distribution for the aggregates in a portfolio~\cite{McNeil:2003}.

Consider a portfolio of $N$ obligors with similar exposure. One possible way to model the entire loss of the portfolio is to calculate
 \begin{align}\label{eq:singlesSevirity}
L_{S}=\sum_{i=1}^N Y_{i}S_{i},
\end{align}
where $Y_{i}$ is a random variable obtained from a Bernoulli distribution which models the default of the $i$-th obligor, and 
$S_{i}$ is a random variable from an arbitrary distribution, so-called severity distribution, modeling the amount of loss of the $i$-th obligor. 
In case of a large portfolio applying Monte Carlo to simulate the loss of each obligor can greatly increase the calculation effort.
This motivates the use of another approach to model the total loss of a given portfolio, namely to consider all $N$ obligors under certain conditions as a single obligor (aggregation process). The total loss is then given by
 \begin{align}\label{eq:compound}
L_{A}=\sum_{i=1}^n X_{i},
\end{align}
where $ n \in  \{1, \cdots ,N, \cdots \}$ is a discrete random variable representing the loss frequency and  $X_i$ are independent and identically distributed random variables representing the loss severity. The distribution of the sum in equation~(\ref{eq:compound}) is called a compound distribution.  A compound distribution is a mix of two distributions. From the first one, called frequency distribution, one obtains an integer $n$, and then generates $n$ random variables using the second distribution, called severity distribution.  In this paper, we compare the two approaches for the calculation of the total loss. To this end, we estimate the quantiles of the single and aggregate loss and study the deviations between  them.

The paper is structured as follows. In section \ref{sec:GammaQuant}, we consider the aggregate loss distribution with a Poisson distribution as a frequency distribution and a Gamma distribution as a severity distribution. Since there is no closed form
for the compound distribution in equation~(\ref{eq:compound}), we introduce
an approximation for the VaR of the aggregate loss. To this end, we introduce a new approach to calculate the quantile function of the Gamma distribution at high confidence levels.
In section \ref{sec:sinleLossApproach}, we consider the single loss distribution.
In section~\ref{sec:exponentialDist},
we apply an exponential distribution as a severity distribution and 
a Bernoulli distribution as a default distribution to calculate the VaR of the single loss distribution and compare the results with those obtained from the aggregation process.
In section~\ref{sec:TrunExponentialDist}, we apply a truncated exponential distribution as the severity distribution and compare the results with those obtained in the previous sections. We conclude our findings in section 4.

\section{Aggregate Loss Distribution}\label{sec:GammaQuant}
In the following, we discuss the distribution of the aggregate loss~(\ref{eq:compound}) for Gamma distributed severities. Analytically,
the compound distribution can be calculated using the method of convolutions \cite{PrilNelson85}.  In case of Gamma distributed severities, the aggregate loss is thus given by the $n$-fold convolution of Gamma distributions, which is a Gamma distribution itself \cite{james2008}. Having calculated the compound distribution, one obtains the VaR at a confidence level $\kappa$, the $\kappa$-quantile, as its inverse function
\begin{equation}
{\rm VaR}_{\kappa}[L_A]=F^{-1}_{L_A}(\kappa) \ .
\end{equation} 
However, there is no closed form for the inverse function of the Gamma distribution.  In section~\ref{sec:GammaEstimateClosed}, we thus discuss a closed-form approximation for the VaR at high confidence levels proposed in \cite{Bocker:2005aa} and use it  to estimate the quantile of the aggregate loss~(\ref{eq:compound}). To this end, we introduce an approximation of the Gamma quantile function in section~\ref{sec:gammaQuantileApp}. 
  
\subsection{Closed-Form Approximation for VaR}\label{sec:GammaEstimateClosed}
Consider  independent and identically distributed severities $X_1, \dots, X_N$ from a heavy-tailed distribution $F$ and a frequency distribution which can be a Poisson, a binomial or a negative binomial distribution. Then, the $\kappa$-quantile of the aggregate loss  $L_A  = X_1 + \cdots + X_N $ satisfies the approximation
 \begin{align}\label{eq:VaRGamma}
{\rm VaR}_{\kappa}[L_A] \rightarrow F^{-1} 
\left(
1- \frac{1- \kappa}{E[N]}
\right),  \;  \text{as}  \;  \;  \kappa \rightarrow 1,
\end{align}
where $E[N] $ is the mean of the frequency distribution. This approximation has been proposed by \cite{Bocker:2005aa} in the context of the Loss Distribution Approach (LDA) to modeling operational risk. 

In our setting, $F$ is the Gamma distribution and the frequency distribution is a Poisson distribution. We evaluate the $\kappa$-quantile of the aggregate loss $L_A$ in equation (\ref{eq:compound}) at the new
confidence level $u$
\begin{equation}\label{eq:XofOurComp}
u:= 1- \frac{1 -\kappa}{E[N]} \ .
\end{equation}
We note that $\kappa $ is the confidence level of $99.5\%$ in Solvency II and $99.9\%$ in Basel III over a capital horizon of one year. Take into account that the closer $u$ to 1 the more precise is the approximation as discussed in~\cite{Bocker:2005aa}. Thus, we can approximate the VaR for the aggregate loss as
\begin{equation}\label{eq:GammaQuantile}
{\rm VaR}_{\kappa}[L_A] =F^{-1}_{L_A}\bigl(\kappa \bigr) \approx 
F_{\Gamma}^{-1}\bigl(u ; \alpha, \beta) \ ,
\end{equation}
where $\alpha$  and $\beta$ are the shape and rate parameter of the Gamma distributed severities.

\subsection{Gamma Quantile Approximation}\label{sec:gammaQuantileApp}
To calculate the VaR of the aggregate loss explicitly, we need the quantile function of the Gamma distribution. There is, however, no closed form for the quantile function of the Gamma distribution; as a result, approximate representations are usually used. 
These approximations generally fall into one of four categories, series expansions, functional approximations, numerical
algorithms or closed form expressions written in terms of a quantile function of another distribution, see e.g. references 
\cite{Steinbrecher:2007Gy, EJM:8678315, Munir2012}.

\begin{figure}[h]
\centering
    \includegraphics[width=0.6\textwidth]{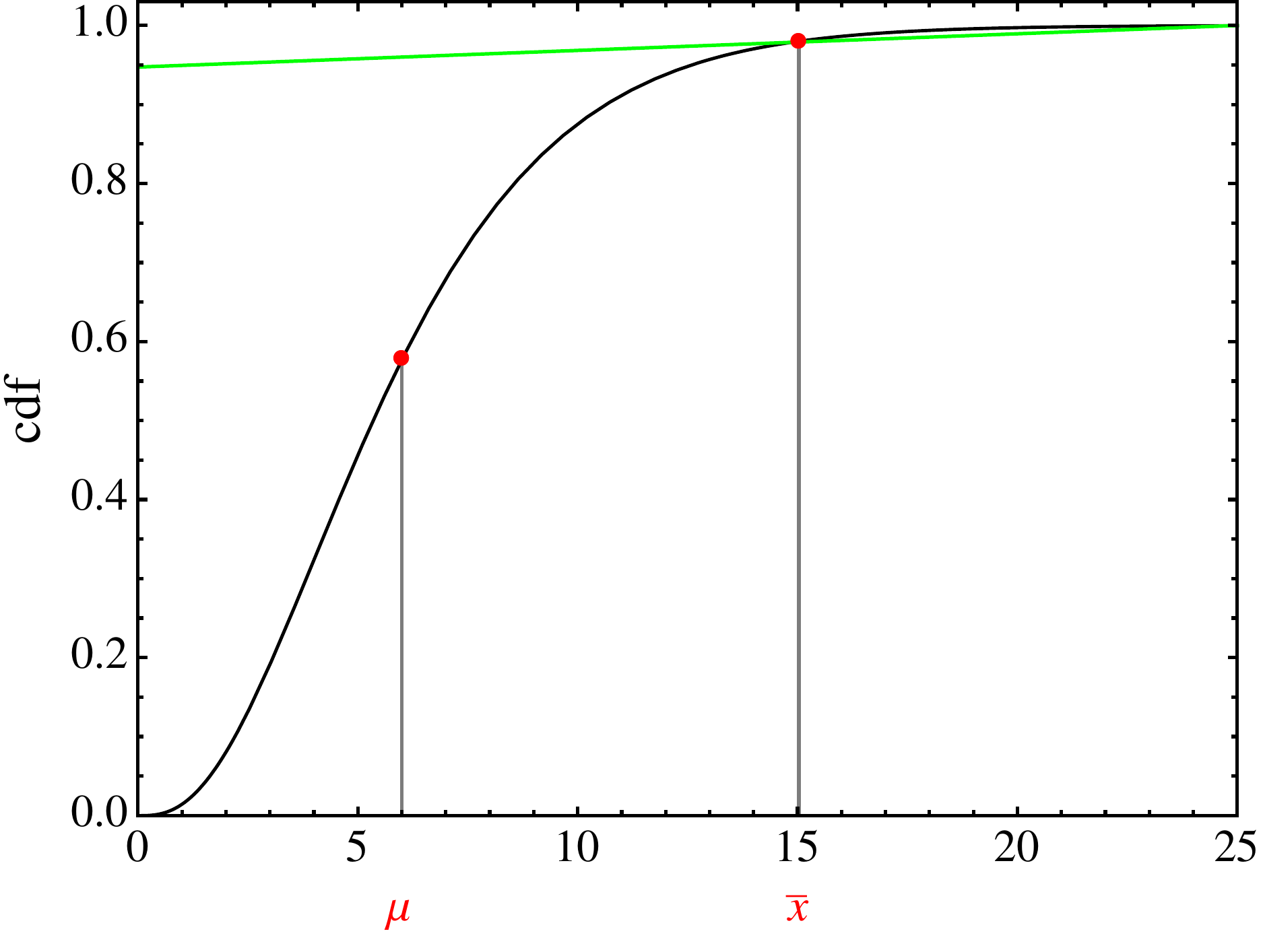}
  \caption{Schematic description of our approach to approximate the tail of the CDF by a linear equation. Here, $\mu$ denotes the mean of the Gamma distribution and $\bar{x}$ is the point at which the quantile function is evaluated. }
  \label{fig1}
\end{figure}

Here, we introduce an approach to estimate the quantile function of the Gamma distribution at high confidence levels. To this end, we consider the tail of the Gamma distribution.
In the tail of the distribution the CDF shows nearly linear behavior, see figure~\ref{fig1}. 
Thus, we can estimate the CDF of the Gamma distribution in the tail by a linear equation
\begin{align}\label{eq:linearCDFAllg}
F_{\Gamma}\bigl(x;\alpha ,\beta \bigr) \approx 
f(\alpha ,\beta ) \; x + y_{\rm int}(\alpha ,\beta )  \ .
\end{align}
$f(\alpha ,\beta ) $  is the slope of the linear equation. It can be calculated as the derivative of the CDF which should be 
evaluated at $\bar{x}= \mu + \Delta $
\begin{align}\label{eq:linearSteigAllg}
f(\alpha ,\beta )|_{x=\bar{x}}= 
\frac{\beta }{\Gamma(\alpha)}
{\rm e}^{-\beta \bar{x}} (\beta \bar{x})^{\alpha -1} \ ,
\end{align}
where $\mu$ denotes the mean of the Gamma distribution, see figure~\ref{fig1}, and $\Delta$ is a shift which we will calculate later on. To find the y-intercept $y_{\rm int}(\alpha,\beta)$, we use the fact that 
the slope of the line should be constant, i.e., we obtain the same slope for the extension of the line to 
the y-axis. As 
\begin{align}\label{eq:linearSlopeAllg}
f(\alpha ,\beta )|_{x=\bar{x}}= 
\frac{F_{\Gamma}\bigl(\bar{x};\alpha ,\beta \bigr)  - y_{\rm int}(\alpha,\beta)}{\bar{x}} \ 
\end{align}
we obtain
\begin{equation}\label{eq:Interception}
y_{\rm int}(\alpha,\beta)=F_{\Gamma}\left(\bar{x};\alpha ,\beta \right)  - {\bar{x}} f(\alpha ,\beta )|_{x=\bar{x}} \ .
\end{equation}
To calculate the quantile function $F^{-1}_{\Gamma}$,
we insert equations~(\ref{eq:linearSteigAllg}) and~(\ref{eq:Interception}) into equation~(\ref{eq:linearCDFAllg}) and invert. This leads to 
\begin{align}\label{eq:lQuantileA}
q_{\Gamma}(u)=F_{\Gamma}^{-1}\left(u; \alpha ,\beta \right)=
\bar{x}+{\rm e}^{\bar{x} \, \beta} \; \bar{x} \; \left(\bar{x}\beta \right)^{-\alpha}
\left( 
u  \Gamma(\alpha) - \gamma(\alpha,\bar{x} \beta ) 
\right),
\end{align}
where 
\begin{align}\label{eq:xBar}
\bar{x} = \mu +\Delta= \frac{\alpha}{\beta} + \frac{\alpha}{\beta} \left(\frac{\gamma(\alpha, \alpha)}
{p(u ,\alpha) \left({\rm e}^{-\alpha} \alpha ^ \alpha + \Gamma(\alpha) \right)} \right) \ .
\end{align}
Note that $\Gamma(\alpha)$ represents the Gamma function and $\gamma(\alpha, z)$ is the incomplete Gamma function defined as
\begin{equation}
\gamma(\alpha, z)=\int\limits_{0}^{z} \ t^{\alpha-1} {\rm e}^{-t} {\rm d} t \  .
\end{equation}
The factor $p(u ,\alpha)$ is a correction factor which depends only on $\alpha$ and $u$. For a fixed $\alpha$ and $u$, it has to be chosen so that the quantile function $q_{\Gamma}(u )$~(\ref{eq:lQuantileA}) reaches a maximum value. We studied the factor  $p(u ,\alpha)$ numerically in the range $1 \leq \alpha \leq 100$
and $0.9 \leq u \leq 0.999$ and found that it can be described by the following expression 
\begin{equation}  
p(u ,\alpha)=a(u) \ {\rm log} \left(b(u) \, \alpha \right) \  ,
\end{equation}
where $a(u)$ and $b(u)$ are polynomial functions of $u$. For further details see appendix A.
In the following, we compare the estimated result $q_{\Gamma}(u)$ with the theoretical Gamma quantile $q_{\Gamma,{\rm th}}(u)$. Table 1 shows the relative error 
\begin{equation}\label{eq:relerror}
\frac{q_{\Gamma}(u)- q_{\Gamma,{\rm th}}(u)}{q_{\Gamma,{\rm th}}(u)} \ 
\end{equation}
for different  $\alpha$, $u=0.95, 0.99, 0.995, 0.999$ and a fixed $\beta=1$. We observe that the deviations are smaller than $1\%$, which illustrates the goodness of the approximation.
 \begin{table}[h]
 \centering
\begin{tabular}{ |ccc || ccc |}
\hline
$u $  & \ $\alpha$  & relative error in $\%$ &$u$  & \ $\alpha$  & relative error in $\%$  \\
\hline
0.95   & \ 1     & -0.01   &0.99& \ 1 & -0.02\\
             &  \ 5       & -0.02  & &\ 5 & -0.08\\
             &  \ 10    & -0.00 & &\ 10 & -0.00\\
             & \ 50    & -0.00  & &\ 50 & -0.00\\
             & \ 100   & -0.00  & &\ 100 & -0.01\\
             & \ 500  & -0.06& &\ 500 & -0.17\\
              & \ 1000   & -0.08 & &\ 1000 & -0.24\\

\hline
\hline
$u $  & \ $\alpha$  & relative error in $\%$ &$u $  & \ $\alpha$  & relative error in $\%$  \\
\hline
0.995   & \ 1     & -0.01   &0.999& \ 1 & -0.10\\
             &  \ 5       & -0.05  & &\ 5 & -0.88\\
             &  \ 10    & -0.00 & &\ 10 & -0.34\\
             & \ 50    & -0.02  & &\ 50 & -0.08\\
             & \ 100   & -0.00  & &\ 100 & -0.15\\
             & \ 500  & -0.18& &\ 500 & -0.53\\
              & \ 1000   & -0.28 & &\ 1000 & -0.63\\
\hline
\end{tabular}
\caption{Relative error (\ref{eq:relerror}) between the approximated (\ref{eq:lQuantileA}) and the theoretical Gamma quantile for different confidence levels. The theoretical Gamma quantile has been obtained with the Mathematica software.}
\label{tab1}
\end{table}

% % %
\section{Single Loss Distribution}\label{sec:sinleLossApproach}
In the following, we discuss the distribution of the single loss~(\ref{eq:singlesSevirity}) for different severity distributions. We derive the compound distribution using the method of convolutions and calculate the $\kappa$-quantile at high confidence levels. In  section~\ref{sec:exponentialDist}, we consider exponential severities and compare the quantile at $99.5\%$ confidence level with the results obtained using the aggregation process. In section~\ref{sec:TrunExponentialDist}, we take severities from a truncated exponential distribution and compare the results with those obtained in the previous sections. 

\subsection{Exponential Severities}\label{sec:exponentialDist}
Consider a portfolio of $N$ obligors. We are interested in the distribution of the total loss
\begin{equation}\label{eq:tl}
L_S  =Y_1 S_1 + \cdots +Y_N S_N
\end{equation}
with severities  $S_n$ drawn from an exponential distribution with the PDF
 \begin{align}\label{eq:pdfExponen}
f_{S}(x; \lambda)= \lambda \, {\rm e}^{-\lambda x} H(x),
\end{align}
where $\lambda >0 $  is the parameter of the distribution and $H$ is  the Heaviside step function with $H(0)=1$. Note that the default of each obligor is modeled by a Bernoulli distribution, which takes the value of 1 if a default occurs and 0 if it does not. Here, we consider the default of all $N$ obligors, i.e., we assume $Y_n =1 \;  \forall \, n=1,\dots,N$. So, we rewrite equation (\ref{eq:tl}) as
 \begin{align}
L_S   = 
 S_1+ \cdots + S_N \ .
\end{align}
Thus, to calculate the distribution of the total loss we have to consider the convolution of $N$ exponential distributions.  For the sake of 
simplicity, we assume that all generating PDFs have the same parameter $\lambda^{\prime}$ given by
\begin{align}\label{eq:LambdaPrimeExp}
\lambda^{\prime} = \; {\rm inf} \{ \lambda_{1}, \cdots \lambda_{N}\} \ .
\end{align}
We obtain the $N$-fold convolution of exponential distributions via a Laplace transformation, which leads to the so-called Erlang distribution
 \begin{align}\label{eq:NconvolutionPDFExp}
f_{L_S}(x; \lambda^{\prime}, N) =f_{S_1} * \cdots * f_{S_N}  = \frac{{\lambda^{\prime}}^{\,n} x ^{N-1}}{(N-1)!}{\rm e}^{-\lambda^{\prime} x} \ .
\end{align}
The corresponding CDF is given by
 \begin{align}\label{eq:CDFErlang}
F_{L_S}(x; \lambda^{\prime}, N) = \int_{0}^{x}f_{L_S}(y;\lambda^{\prime}, N) \ dy =
   \frac{\gamma (N,\lambda^{\prime} x)}{(N-1)!}  \ .
\end{align}
We note the resemblance with the CDF of the Gamma distribution
\begin{align}\label{eq:CDFofGamma}
F_{\Gamma}(x)=\frac{\gamma(\alpha,\beta x)}{\Gamma(\alpha)}=\frac{\gamma(\alpha,\beta x)}{(\alpha-1)!}\;
, \; \; \; 
\text{for} \; \alpha>0  \; \text{and} \; \beta>0 \ ,
\end{align}
where we identify $N$ with $\alpha$ and $\lambda^{\prime}$ with $\beta$. This allows us to apply the previous result (\ref{eq:lQuantileA}) to calculate the quantile function of $F_{L_S}(x; \lambda^{\prime}, N)$ at high confidence levels
\begin{align}\label{eq:lQuantileExpA}
q_{L_S}(\kappa)=F_{L_S}^{-1}\bigl(\kappa ; \lambda^{\prime}, N\bigr)=
\bar{x}+{\rm e}^{\bar{x} \, \lambda' } \; \bar{x} \; \bigl(\bar{x}\lambda'  \bigr)^{-N}
\Bigl( 
 \kappa  \Gamma(N) - \gamma(N, \lambda^{\prime} \bar{x} )  
\Bigr) \ .
\end{align}
We now compare the quantile of the single loss distribution at $99.5\%$ confidence level with the quantile of the aggregate loss
\begin{align}\label{eq:lQuantileGammaA}
q_{L_A}(u)=F_{L_A}^{-1}\bigl(u ;\alpha' ,\beta \bigr)=
\bar{x}+{\rm e}^{\bar{x} \, \beta } \; \bar{x} \; \bigl(\bar{x}\beta \bigr)^{-\alpha'}
\Bigl( 
 u  \Gamma(\alpha') - \gamma(\alpha',\bar{x} \beta ) 
\Bigr)  \ .
\end{align}  
We recall the relation $u= 1- (1-\kappa)/E[N] $ and set $\alpha' = (N+\sigma_{Poisson}) \alpha = ( N+\sqrt{N}) \alpha$. Note that the parameters $\alpha$, $\beta$ and $\lambda^{\prime}$ have to be chosen in a way that the mean of the Erlang and the Gamma distributions 
are equal, i.e. $\mu:=\alpha/\beta = 1/ \lambda^{\prime} $.  In the following, we study the difference between the quantiles $q_{L_S}(\kappa)$ and $q_{L_A}(u)$
\begin{equation}\label{eq:diffOFquantA}
d_{S,A}(\kappa; N)=
q_{L_S}(\kappa)-q_{L_A}(u) \ .
\end{equation}
Figure 2 shows the absolute $d_{S,A}(\kappa; N)$ and the relative difference $d_{S,A}(\kappa; N)/q_{L_S}(\kappa)$ between the quantiles. Although the absolute difference is increasing as the mean and the number of obligors $N$ grow, the relative difference decreases. 
\begin{figure}[h]   
  \begin{subfigure}[b]{0.495\textwidth}
    \includegraphics[width=\textwidth]{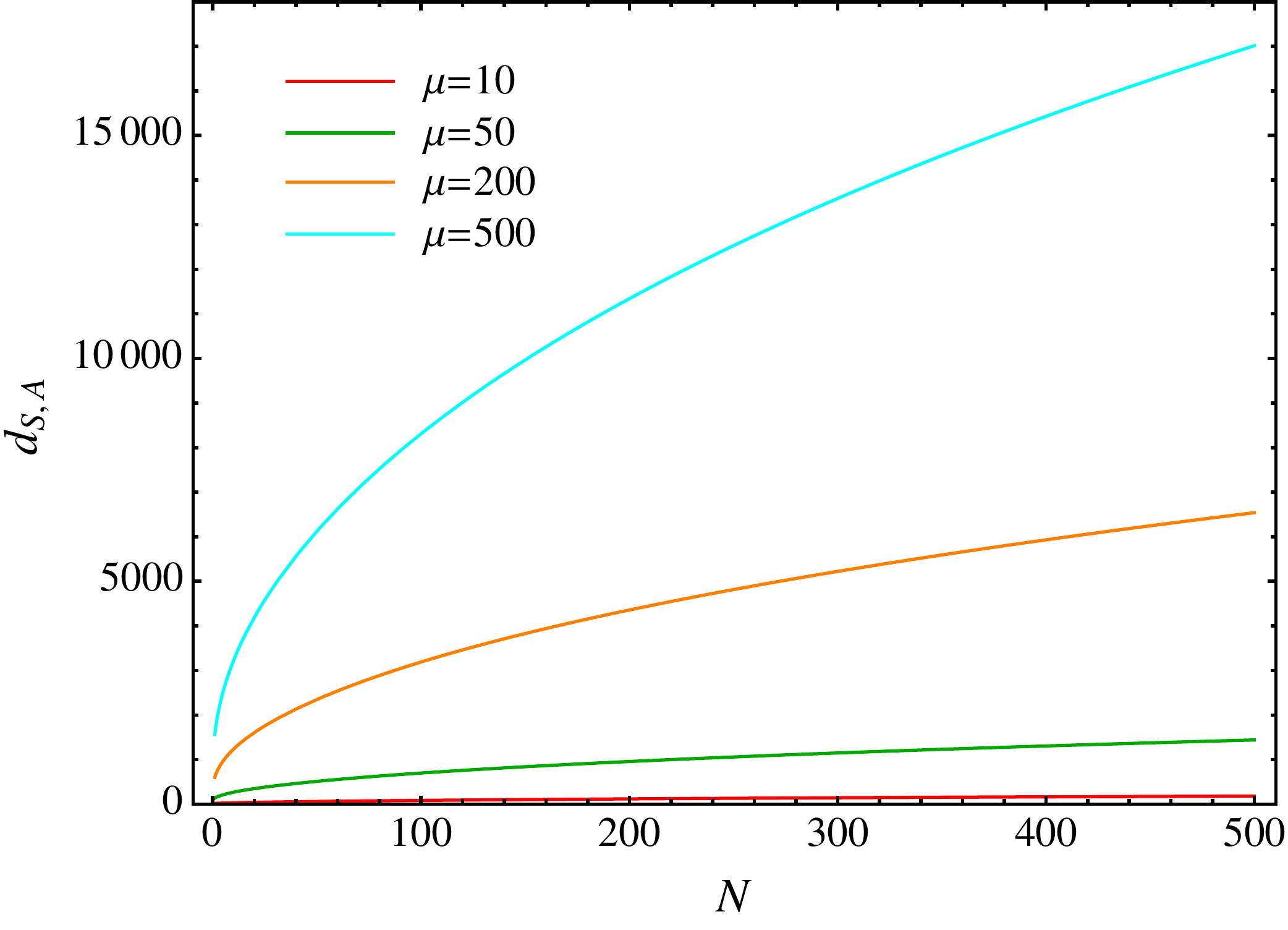}
    \caption{Absolute difference between $q_{L_S}(\kappa)$ and $q_{L_A}(u)$.}
    \label{fig:absolutediff}
  \end{subfigure}
  \hfill
  \begin{subfigure}[b]{0.48\textwidth}
    \includegraphics[width=\textwidth]{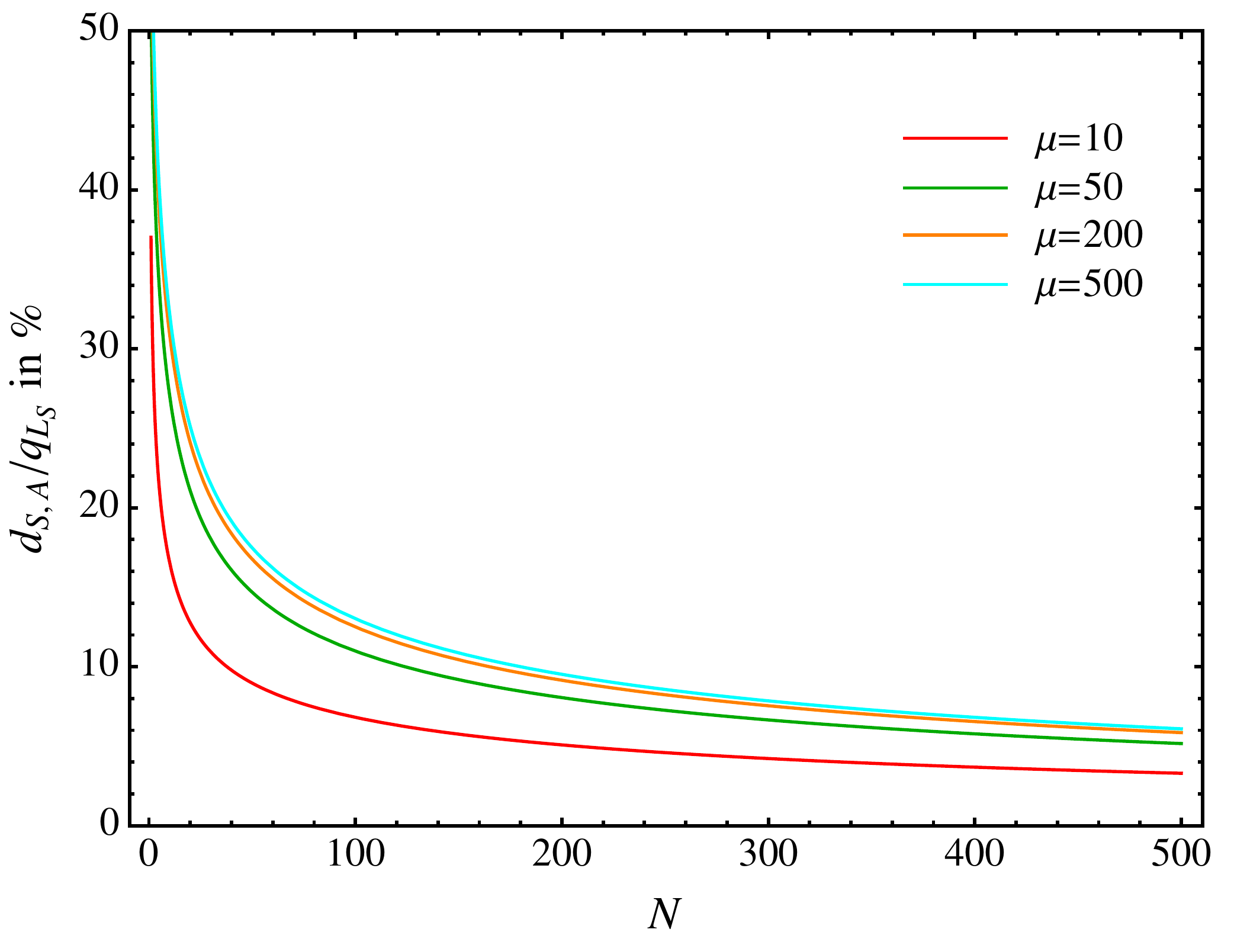}
    \caption{Relative difference between $q_{L_S}(\kappa)$ and $q_{L_A}(u)$. }
    \label{fig:relative2}
  \end{subfigure}
  \caption{Comparison of the quantile of the single loss distribution with exponential severities $q_{L_S}(\kappa)$ (\ref{eq:lQuantileExpA}) and the quantile of the aggregate loss with Gamma distributed severities $q_{L_A}(u)$ (\ref{eq:lQuantileGammaA}) at $\kappa=0.995$. }
  \label{fig:diffExp}
\end{figure}
This is due to the fact that the absolute value of the quantile $q_{L_S}(\kappa)$ is growing faster than the absolute difference. Furthermore,  we observe that the  relative difference shows a convergent behavior for high $N$ values which is independent of the value of the mean $\mu$. 

\subsection{Truncated Exponential Severities}\label{sec:TrunExponentialDist}
In practice, the total loss cannot be greater than the gross exposure $L$, i.e., the maximum possible loss.  Therefore, a truncated distribution is often used to cut the loss distribution at  the gross exposure. A truncated distribution is a conditional distribution obtained by restricting the domain of some other probability distribution \cite{RePEc2015}.  Here, we apply a truncated exponential distribution to model the severities $S_n$ of the total loss $L_S$ (\ref{eq:tl}). The corresponding PDF reads
\begin{align}\label{eq:pdfOfTexp}
f _{S | 0< S < L}\left(x;\lambda,L\right)= \frac{\lambda {\rm e}^{-\lambda x}}{1-{\rm e}^{-\lambda L}} \ ,
\end{align}
where $L>0$ denotes the gross exposure. As in the previous section we assume  $Y_n =1 \;  \forall \, n=1,\dots,N$. Thus, to calculate the loss distribution we need the convolution of $N$ truncated exponential distributions with $\lambda^{\prime} = \; {\rm inf} \{ \lambda_{1}, \cdots \lambda_{N}\}$. Again, applying a Laplace transformation, we obtain
\begin{align}\label{eq:nTimesConv}
f_{L_S}(x;\lambda^{\prime},N, L)=f_{S_1}\ast \cdots \ast f_{S_N}=
\frac{{\rm e}^{-\lambda^{\prime}x} {\lambda^{\prime}}{N}x^{N-1}}
{(N-1)! \bigl(1-  {\rm e}^{-L \lambda^{\prime}}\bigr) ^N} \ .
\end{align}
Then, the corresponding CDF reads 
\begin{align}\label{eq:CDFnTimesConv}
F_{L_S}(x; \lambda^{\prime}, N, L )&= 
\int_{0}^{x}f_{L_S}(y;\lambda^{\prime},N, L)\, dy  \notag \\
&=\frac{\bigl( 1-{\rm e}^{ -\lambda^{\prime}L}\bigr)^{-N}}
{\Gamma(N)}
\Bigl(
\Gamma(N) - \Gamma(N, \lambda^{\prime}x)
\Bigr)\notag \\
&=
\frac{1}{\bigl( 1-{\rm e}^{- \lambda^{\prime}L}\bigr)^{N}}
\frac{\gamma(N,\lambda^{\prime}x)}
{\Gamma(N)}  \ ,
\end{align}
where in the last step we used the relation $\gamma( N,\lambda^{\prime}x) + \Gamma(N, \lambda^{\prime}x)=\Gamma(N) $ between the incomplete  $\gamma( N,\lambda^{\prime}x)$ and the upper Gamma function  $\Gamma(N, \lambda^{\prime}x)$. Again, we recognize the resemblance with the CDF of the Gamma distribution up to the factor $1/\bigl( 1-{\rm e}^{- \lambda^{\prime}L}\bigr)^{N}$. Applying our result (\ref{eq:lQuantileA}), we obtain the quantile function
\begin{align}\label{eq:lQuantileTrunExpA}
q_{L_S}(\kappa')=F_{L_S}^{-1}\bigl(\kappa' ; \lambda^{\prime}, N, L \bigr)=
\bar{x}+{\rm e}^{\bar{x} \, \lambda^{\prime}} \; \bar{x} \; \bigl(\lambda ^{\prime} \bar{x} \bigr)^{-N}
\Bigl( 
 \kappa'  \; \Gamma(N) - \gamma(N, \lambda^{\prime}  \bar{x}) 
\Bigr) \ ,
\end{align}
where 
\begin{align}\label{eq:lQuantileKappa}
\kappa^{\prime} = \bigl(
 1-{\rm e}^{- \lambda^{\prime}L} \bigr)^{N} 
 \kappa \ .
 \end{align}
 As  in the previous section, we are interested in the difference between the quantiles $q_{L_S}(\kappa^{\prime})$ and $q_{L_A}(u)$ 
 \begin{equation}\label{eq:diffOFquantKappa}
d_{S,A}(\kappa^{\prime}; N)=
q_{L_S}(\kappa^{\prime})-q_{L_A}(u) \ .
\end{equation}
It is important that we have to choose the parameters $\lambda^{\prime}$, $L$, $\alpha$ and $\beta$ in a way
that the mean of truncated exponential distribution is equal to the mean of the Gamma distribution, i.e.,
\begin{align}\label{eq:MenCondition}
\mu=\frac{1-e^{L \lambda^{\prime} }+ L\lambda^{\prime}}{\lambda^{\prime} ( 1- e^{L \lambda^{\prime}})}
= 
\frac{\alpha}{\beta} \ .
\end{align}
Usually, the gross loss  $L$ is fixed, i.e., we have to set $\lambda^{\prime}$ and $\mu$.  We assume that 
\begin{equation}\label{eq:multipleOfL}
L = \mathcal{C} \frac{1}{\lambda^{\prime}} , \; \;
\text{where} \quad \mathcal{C} \in \mathbb{R} \  .
\end{equation}

\begin{figure}[tp]
\centering
    \includegraphics[width=0.5\textwidth]{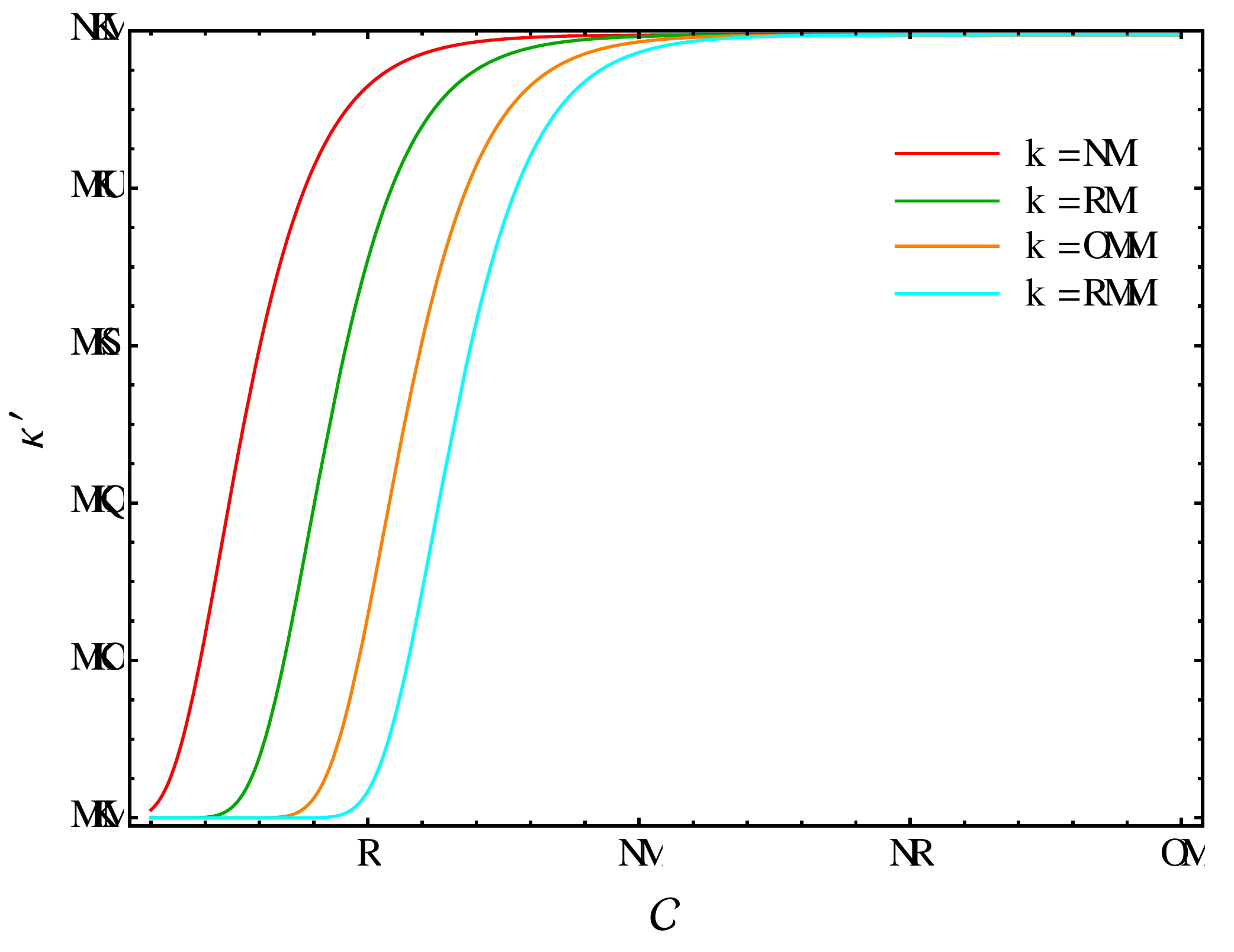}
\caption{ Dependence between $ \kappa^{\prime}$ (\ref{eq:lQuantileKappa}) and the constant $\mathcal{C}$ (\ref{eq:multipleOfL}).}
  \label{fig:kappaTexp}
  \end{figure}

The gross loss can be viewed as a multiple of the mean $1/\lambda^{\prime}$ of the underlying exponential distribution. Then, the constant $ \mathcal{C}$ results from the chosen model. Note that $ \mathcal{C}$ occurs 
in the power of the exponential function in equation (\ref{eq:lQuantileKappa})  and plays an important role for the determination of the quantile function.
We can see in figure~\ref{fig:kappaTexp} that $\kappa^{\prime} $ depends mainly on $\mathcal{C} $ but also on $N$.
Since we determined equation (\ref{eq:lQuantileA}) for the range  $0.9 \leq u \leq 0.999$ we have to take into account 
that $ \mathcal{C} $ should be greater than 9 in our consideration.
  
\begin{figure}[hp]   
  \begin{subfigure}[b]{0.495\textwidth}
    \includegraphics[width=\textwidth]{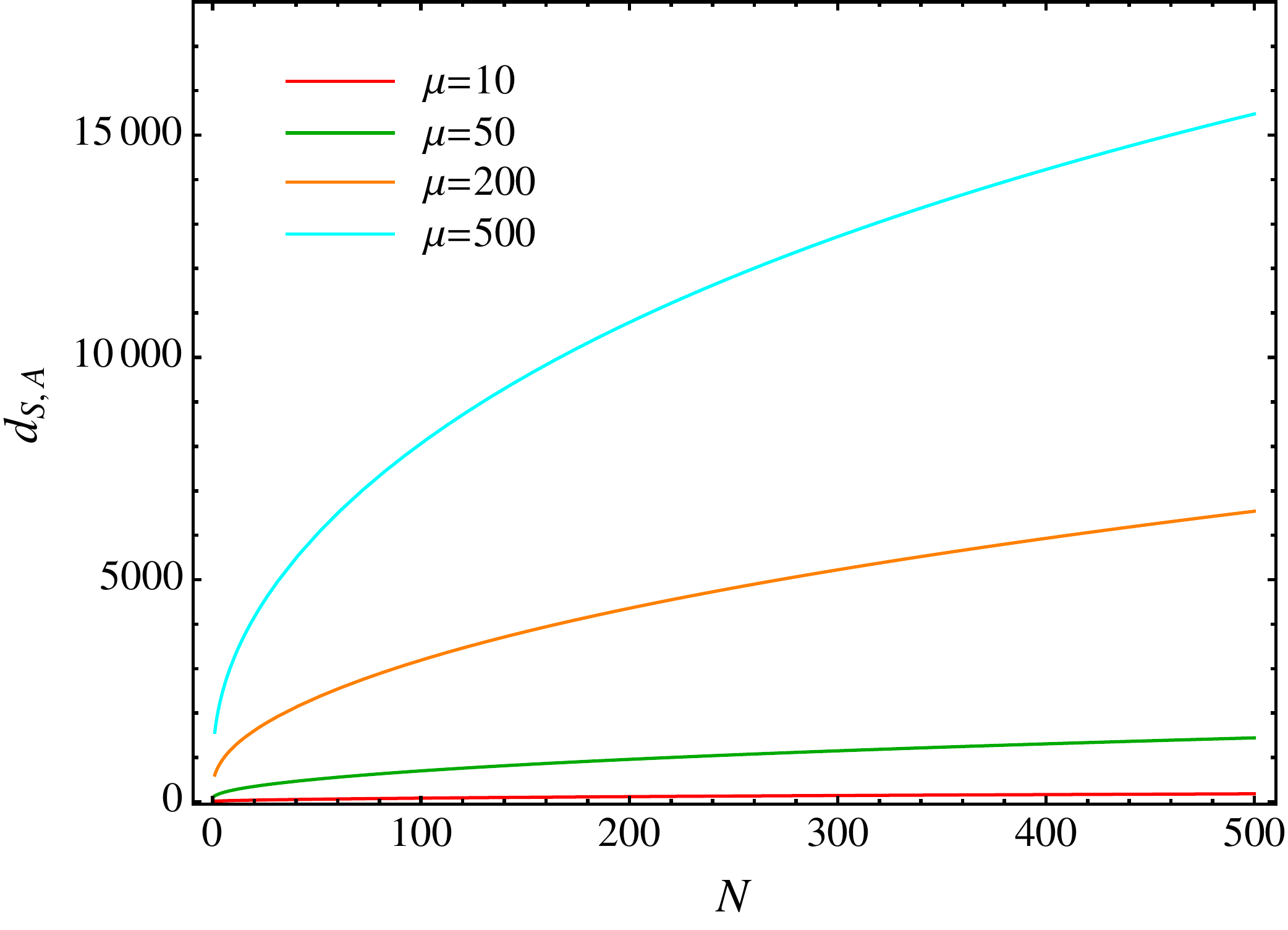}
    \caption{Absolute difference between $q_{L_S}(\kappa^{\prime})$ and $q_{L_A}(u)$.}
    \label{fig:absolutediffTexp6}
  \end{subfigure}
  \hfill
  \begin{subfigure}[b]{0.48\textwidth}
    \includegraphics[width=\textwidth]{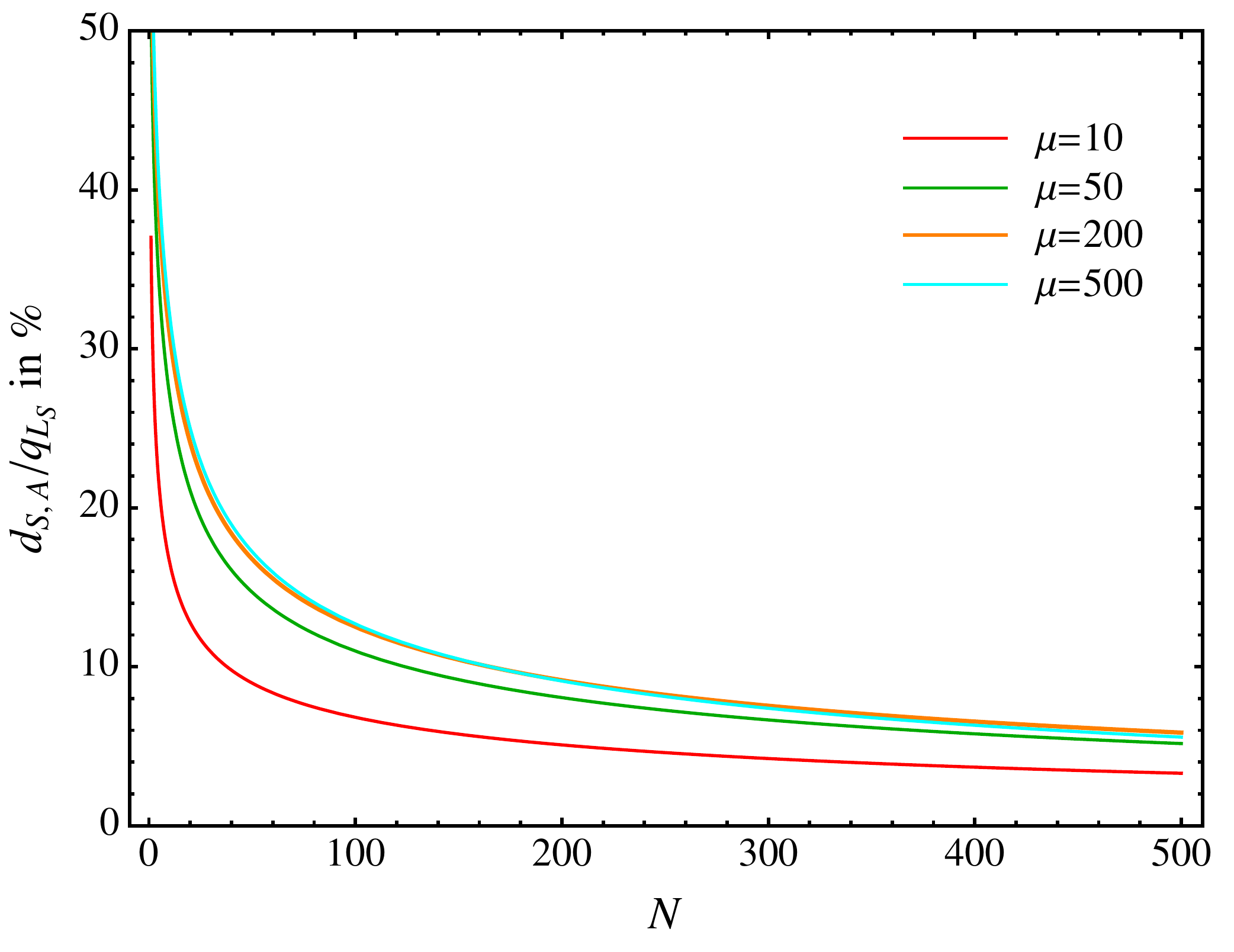}
    \caption{Relative difference  between $q_{L_S}(\kappa^{\prime})$ and $q_{L_A}(u)$.}
    \label{fig:relativeTex6}
  \end{subfigure}
  \caption{Comparison of the quantile of the single loss distribution with truncated exponential severities $q_{L_S}(\kappa^{\prime})$ (\ref{eq:lQuantileTrunExpA}) for $L=6000$ and the quantile of the aggregate loss with Gamma distributed severities $q_{L_A}(u)$ (\ref{eq:lQuantileGammaA}) at $\kappa=0.995$.}
  \label{fig:diffTExp6}
\end{figure}

\begin{figure}[h]   
  \begin{subfigure}[b]{0.495\textwidth}
    \includegraphics[width=\textwidth]{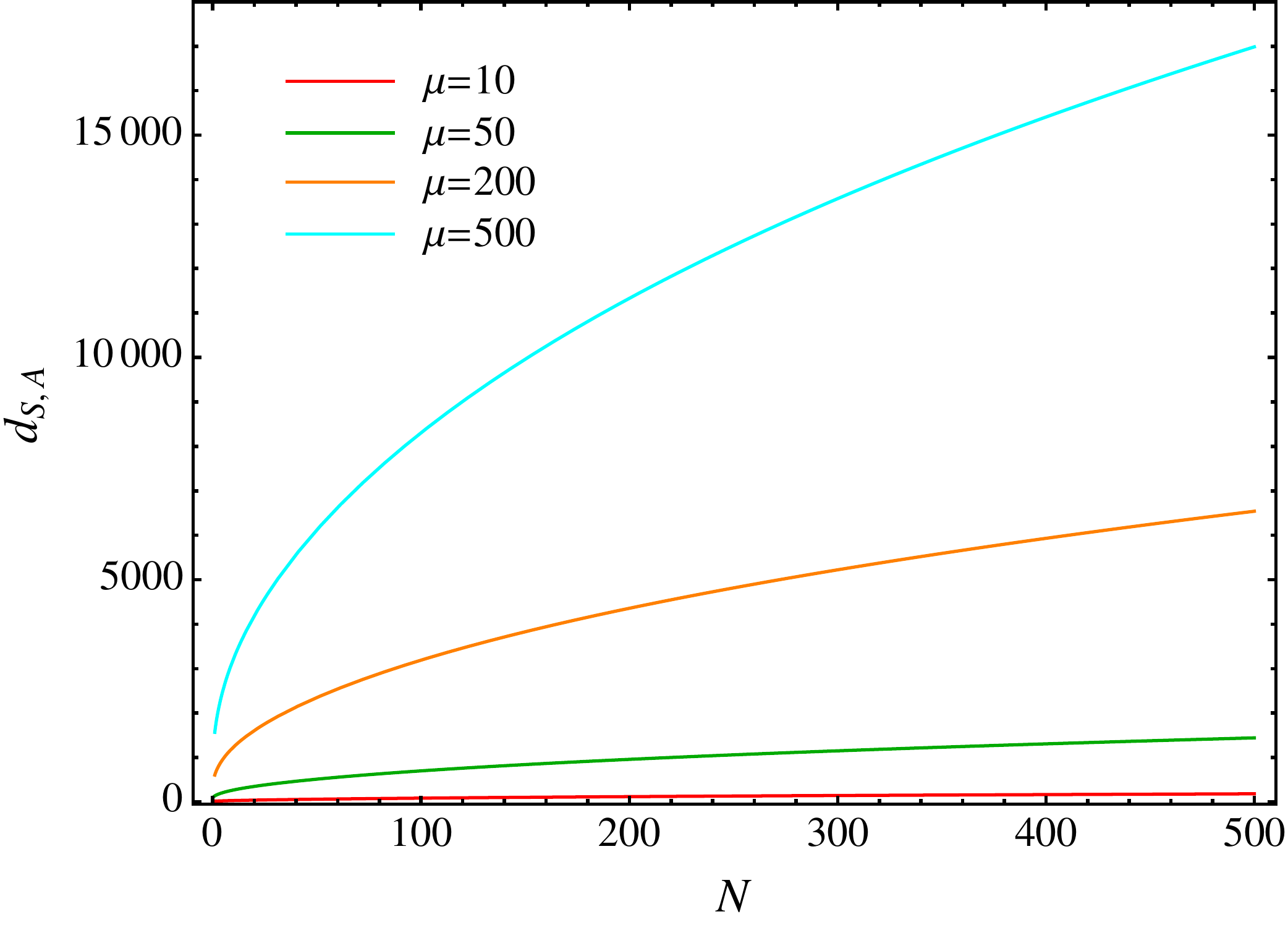}
    \caption{Absolute difference between $q_{L_S}(\kappa^{\prime})$ and $q_{L_A}(u)$.}
    \label{fig:absolutediffTexp}
  \end{subfigure}
  \hfill
  \begin{subfigure}[b]{0.48\textwidth}
    \includegraphics[width=\textwidth]{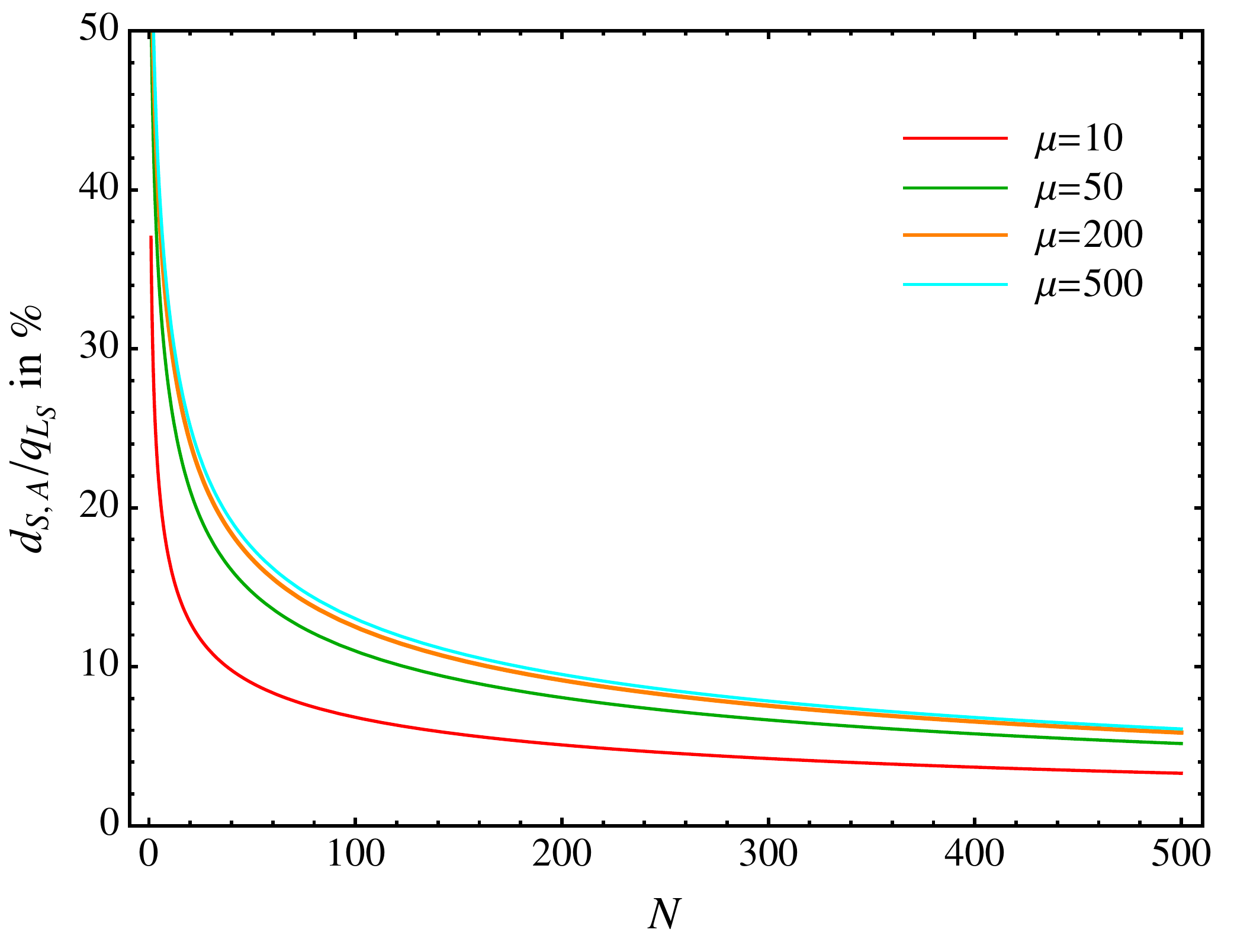}
    \caption{Relative difference between $q_{L_S}(\kappa^{\prime})$ and $q_{L_A}(u)$. }
    \label{fig:relativeTex}
  \end{subfigure}
  \caption{Comparison of the quantile of the single loss distribution with truncated exponential severities $q_{L_S}(\kappa^{\prime})$ (\ref{eq:lQuantileTrunExpA}) for $L=8000$ and the quantile of the aggregate loss with Gamma distributed severities $q_{L_A}(u)$ (\ref{eq:lQuantileGammaA}) at $\kappa=0.995$.}
  \label{fig:diffTExp}
\end{figure}

In the following, we study the absolute as well as the relative difference between $q_{L_S}(\kappa^{\prime})$ and $q_{L_A}(u)$ (\ref{eq:lQuantileGammaA}) for two different values of the gross exposure $L=6000$ and $L=8000$, see figures \ref{fig:diffTExp6} and \ref{fig:diffTExp}, respectively. To this end, we set $\kappa = 0.995$ and $\alpha' = (N+\sigma_{Poisson}) \alpha = ( N+\sqrt{N}) \alpha$ in equation
(\ref{eq:lQuantileGammaA}) and determine $\lambda^{\prime}$ from equation (\ref{eq:MenCondition}) for a fixed $\mu$.
% L= 8000
We observe that both the absolute and the relative difference decrease the lower the gross loss $L$ becomes. 
An interesting observation in figure \ref{fig:relativeTex6} is that the relative difference  for $\mu = 500$  falls quicker than the one for $\mu = 200$.
The reason is that $d_{S, A}(\kappa^{\prime}; N)$ for $\mu = 500$ approaches  the limit $L= 6000$ quicker and decreases quicker.
To illustrate this the values of $d_{S,A}(\kappa^{\prime}; N)$ and $d_{S,A}(\kappa^{\prime}; N)/q_{L_S}(\kappa^{\prime})$  are shown in table \ref{tab:comparisonTab} for both $L=6000$ and $L=8000$ with $N=500$ and $\mu =500$. According to equation (\ref{eq:lQuantileKappa}) the confidence level is shifted from $\kappa=0.995$ to $\kappa^{\prime}$. In addition, table \ref{tab:comparisonTab} shows a comparison with the case of  the exponential severities  discussed in section~\ref{sec:exponentialDist}. As $L$ grows the truncated exponential model approaches the results obtained in the simple exponential model. However, the truncated exponential distribution as a model for the severities yields results closer to those obtained from the aggregation process. 

\begin{table}[h]
\centering
%\tbl{Exponential vs truncated exponential severities.}
\begin{tabular}{@{}|ll|lll|}
\hline
\multicolumn{2}{|c|}{Exponential severities} & \multicolumn{3}{c|}{\begin{tabular}[c]{@{}c@{}}Truncated exponential  severities\end{tabular}} \\ \hline 
                      & $\kappa=0.995$                   &                  & $\kappa^{\prime}=0.991$              & $\kappa^{\prime}=0.994$            \\[0.05cm]
                      &                      &                  & $L=6000$             & $L=8000$            \\  [0.3cm]
$d_{S,A}(\kappa; N)$                     &  17013.22                     & $d_{S,A}(\kappa^{\prime}; N) $              &      15475.54            &     16985.01           \\[0.05cm]
$d_{S,A}(\kappa; N)/q_{L_S}(\kappa)$                     &    6.09                   & $d_{S,A}(\kappa^{\prime}; N)/q_{L_S}(\kappa^{\prime})$               &  5.57                 &       6.08           \\ \hline
\end{tabular}
\caption{Exponential vs truncated exponential severities.}
\label{tab:comparisonTab}
\end{table}

\section{Conclusion}
We considered an aggregation process, where obligors in a huge portfolio are put together under certain conditions and considered as a single obligor, and estimated the VaR for Gamma distributed severities at high confidence levels.  To this end, we introduced an approach for the semi-analytical calculation of the quantile function of the Gamma distribution and derived an expression which showed a good approximation to the theoretical Gamma quantile function at high confidence levels. 

In addition, we calculated the VaR for a single loss distribution where the severities are drawn first from an exponential and then from a truncated exponential distribution. To this end, we used the method of convolutions and derived an expression for the VaR in both cases. 
We compared the VaR for the single loss distribution with the VaR for the aggregation process and studied the difference between both quantiles.
The relative difference depends on the number of obligors in the portfolio, but also on the amount of gross loss in case of truncated exponential severities. We observe that the truncated exponential distribution as a model for the severities yields results closer to those obtained from the aggregation process.

\appendix
\section{Determination of the correction factor ${p(u,\alpha)}$}

Here, we present some details of the determination of the correction factor ${p(u,\alpha)}$ in equation~(\ref{eq:xBar}).

\begin{figure}[h!]
\centering
    \includegraphics[width=0.6\textwidth]{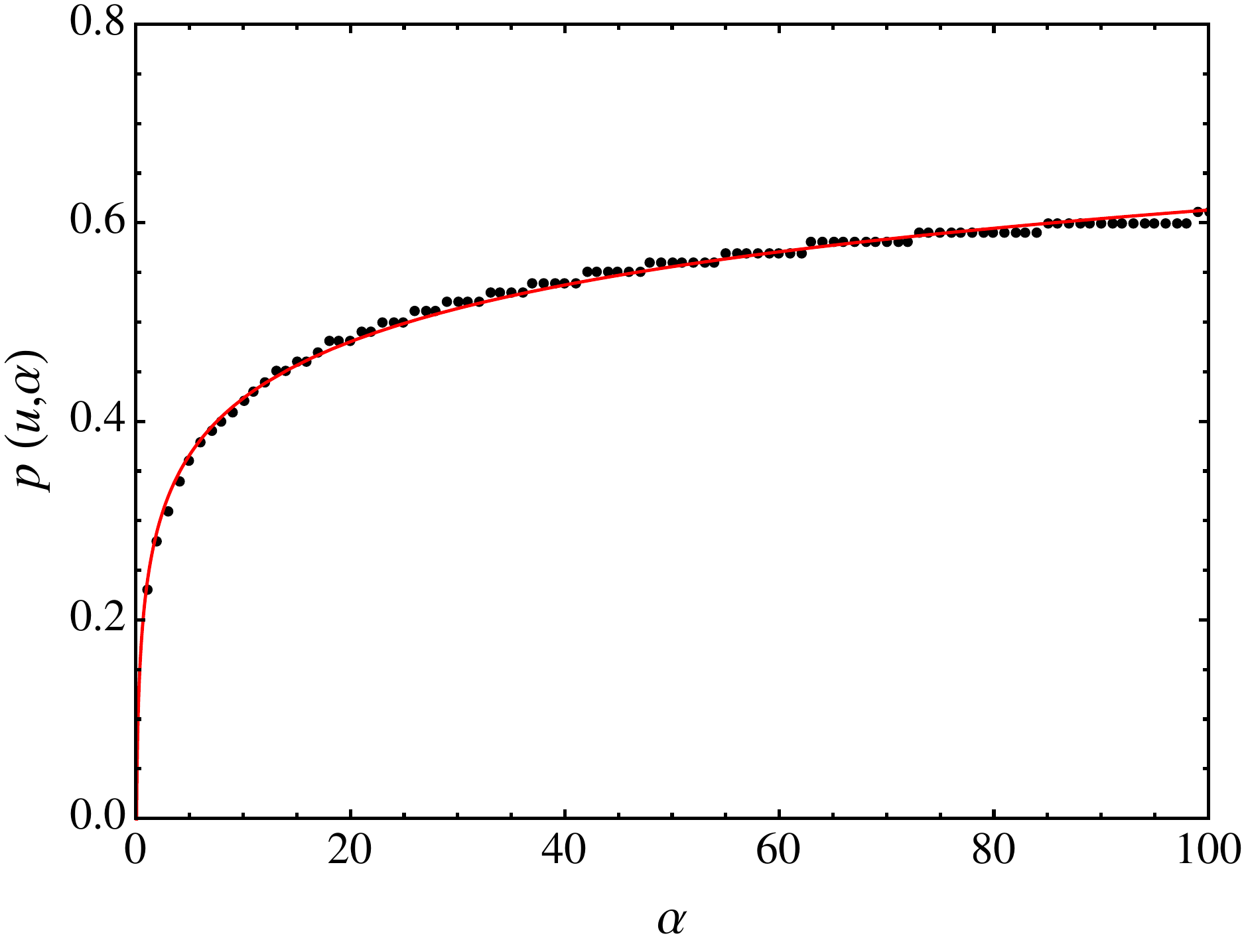}
  \caption{Dependence of the correction factor $p(u,\alpha)$ on $\alpha$ for a fixed $u=0.95$ and $\beta=1$. The red curve represents a log-fit of the form~(\ref{eq:fit1}) with constants $a=0.082$ and $b=17.007$.}
  \label{ap_fig1}
  \end{figure}

\begin{figure}[h!]
\centering
    \includegraphics[width=\textwidth]{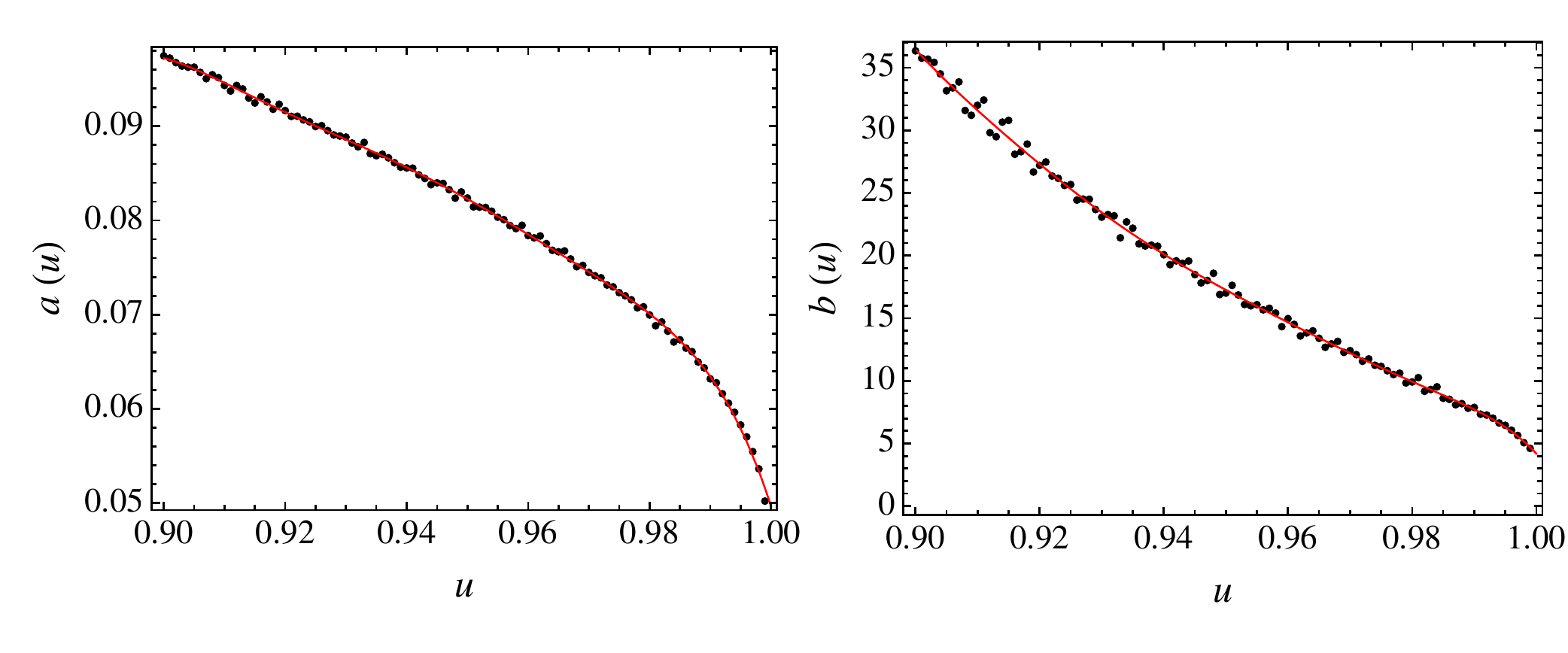}
  \caption{Left: Dependence of $a(u)$ on $u$. The red curve represents a polynomial fit of the form~(\ref{eq:pf1}) with constants $c_0=-4.83 \times 10^5, c_1=3.08 \times 10^6, c_2=-8.16 \times 10^6, c_3=1.16 \times 10^7, c_4=-9.19 \times 10^6, c_5=3.90 \times 10^6, c_6=-6.90 \times 10^5$. Right: Dependence of $b(u)$ on $u$. The red curve represents a polynomial fit of the form~(\ref{eq:pf2}) with constants $d_0=4.35 \times 10^9, d_1=-3.23 \times 10^{10}, d_2=1.02 \times 10^{11}, d_3=-1.80 \times 10^{11}, d_4=1.91 \times 10^{11}, d_5=-1.21 \times 10^{11}, d_6=4.26 \times 10^{10}, d_7=-6.44 \times 10^9$. }
 \label{ap_fig2}
\end{figure}

We study the quantile function~(\ref{eq:lQuantileA}) numerically in the range $1 \leq \alpha \leq 100$
and $0.9 \leq u \leq 0.999$ by varying the correction factor between $0.05$ and $1.5$ and determine the correction factor which maximizes the quantile function. For a fixed $u$, we observe that the correction factor grows as a function of $\alpha$, see figure~\ref{ap_fig1}. The dependence can be described by a log-function of the form
\begin{equation}\label{eq:fit1}
p(u, \alpha)= a \, {\rm log}(b \, \alpha)
\end{equation}
with constants $a$ and $b$. 
Note that the value of the constants depends on the chosen $u$. In the range $0.9 \leq u \leq 0.999$, we observe a decreasing trend for both constants, see figure~\ref{ap_fig2}. 
This behavior can be approximated by polynomial expressions of the form
\begin{align}
a(u)&=  \sum_{i=0}^6 c_i u^i  \ , \label{eq:pf1} \\
b(u)&=  \sum_{i=0}^7 d_i u^i \label{eq:pf2} \ .
\end{align}
Thus, we finally obtain 
\begin{equation}  
p(u,\alpha)=a(u) \ {\rm log} \left(b(u) \, \alpha \right) \  .
\end{equation}
Note that  the precision of the quantile function~(\ref{eq:lQuantileA}) depends highly on the precision of the fit functions and the considered ranges of the parameters $\alpha$ and $u$.

%%%%%%%%%%%%%%%%%%%%%%%%%%%%%%%      Bibliography
\newpage

\bibliographystyle{halpha}
%\bibliography{biblio}

\end{document}